%% file: main.tex
\documentclass[conference,top=1.9cm]{IEEEtran}

\usepackage{xcolor}
\usepackage{balance}

\usepackage{amsmath,amssymb,amsfonts}

\usepackage{graphicx}
\usepackage{subcaption}
\usepackage{textcomp}
\usepackage{xcolor}
\usepackage[colorlinks=false, hidelinks]{hyperref}
\usepackage{cleveref}
\usepackage{siunitx}

\usepackage{tikz}
\usepackage{textcomp}

\usepackage{bm}

\usepackage[ruled,norelsize]{algorithm2e}

\usepackage[
backend=bibtex,
natbib=true,
url=false, 
doi=true,
eprint=false,
sorting=none,
maxbibnames=99,
defernumbers=true,
style=ieee,
giveninits=true,
]{biblatex}
\AtEveryBibitem{
 \clearlist{address}
 \clearfield{url}
 \clearfield{month}
 \clearfield{editor}
 \clearfield{eprint}
 \clearfield{volume}
 \clearfield{isbn}
 \clearfield{issn}
 \clearlist{location}
 \clearlist{editor}
 \clearfield{note}
 \clearfield{pages}
}
\newcommand\copyrighttext{%
  \footnotesize \textcopyright 2025 IEEE. Personal use of this material is permitted.
  Permission from IEEE must be obtained for all other uses, in any current or future 
  media, including reprinting/republishing this material for advertising or promotional 
  purposes, creating new collective works, for resale or redistribution to servers or 
  lists, or reuse of any copyrighted component of this work in other works. 
  }
\newcommand\copyrightnotice{%
\begin{tikzpicture}[remember picture,overlay]
\node[anchor=south,yshift=10pt] at (current page.south) {\fbox{\parbox{\dimexpr\textwidth-\fboxsep-\fboxrule\relax}{\copyrighttext}}};
\end{tikzpicture}%
}

\usepackage{glossaries}
\def\BibTeX{{\rm B\kern-.05em{\sc i\kern-.025em b}\kern-.08em
    T\kern-.1667em\lower.7ex\hbox{E}\kern-.125emX}}

\input{abbrevs}
\input{macros}

\begin{document}

\title{Joint Delay-Doppler Estimation using OFDMA Payloads for Integrated Sensing and Communications}

\author{\IEEEauthorblockN{
Marc Miranda\IEEEauthorrefmark{1},   
Sebastian Semper\IEEEauthorrefmark{1}\IEEEauthorrefmark{2},   
Christian Schneider\IEEEauthorrefmark{1}\IEEEauthorrefmark{2},    
Reiner Thom\"a\IEEEauthorrefmark{1},      
Giovanni Del Galdo\IEEEauthorrefmark{1}\IEEEauthorrefmark{2}
}

\IEEEauthorblockA{
\IEEEauthorrefmark{1}
Technische Universität Ilmenau, Ilmenau, Germany, {marc}.{miranda}@tu-ilmenau.de}
\IEEEauthorrefmark{2}
Fraunhofer Institute for Integrated Circuits IIS, Ilmenau, Germany
}

\maketitle

\copyrightnotice

\begin{abstract}
The use of future communication systems for sensing offers the potential for a number of new applications. In this paper, we show that leveraging user data payloads in multi-node Orthogonal Frequency Division Multiple Access (OFDMA) networks for estimating target delay and Doppler-shift parameters can yield a significant advantage in SNR and addressable bandwidth. However, gaps in the frequency-time resources, reference signal boosting and amplitude modulation schemes introduce challenges for estimation at the sensing receiver. 

In this work, we propose a joint delay and Doppler-shift model-based estimator designed to address these challenges. Furthermore, we demonstrate that incorporating knowledge of the device model into the estimation procedure helps mitigate the effects of the non-ideal radar ambiguity function caused by amplitude-modulated user payloads and sparse reference signals. Simulation results demonstrate that the estimator achieves the theoretical lower bound on estimation variance.
\end{abstract}

\vskip0.5\baselineskip

\section{Introduction}
\gls{icas} is considered one of the key features of future 6G mobile communication \cite{isac_springer_2023}, \cite{jsrs_thoma_dallman_2021}, \cite{jrc_arch_shatov_2024}. In essence, \gls{icas} is a means of radar detection and localization of passive objects ("targets") that are not equipped with a radio tag. The radio signals transmitted by the mobile radio nodes for communication purposes are reused for target illumination. At the same time, the mobile communication network ensures data fusion and processing. This is most resource efficient since the resources provided by the existing mobile radio network are simultaneously used for two services, communication and sensing. Moreover, mobile communication already has a wealth of orthogonal access and radio link control schemes included that can be used for advanced multi-sensor access, coordination and link adaptation \cite{distributed_icas_thomae_2023}. However, because of the random nature and modulation schemes of signals used for communications in the 5G NR, it seems obstructive to use the communication payload for sensing purposes since the waveforms are “not designed for sensing” \cite{isac_vehicle_nataraja_2024}. Therefore, it is often proposed to use well defined reference signals (CSI, DM and PRS) for radar sensing. However, using only a smaller part of the transmitted signal for sensing reduces sensing sensitivity and increases the sparsity of the available samples in time and frequency, resulting in non-ideal conditions for parameter estimation. 

Using the full (user dependent) data payload for \gls{icas} was already proposed in \cite{cpcl_thoma_2019} as \gls{cpcl} which is a reminiscent of passive radar. In \cite{prob_const_shaping_geiger_2025}, a modulation constellation shaping approach was studied to improve the detection probability when using random user data for sensing. Similarly, related issues when using LTE signals for passive radar have been outlined in \cite{ref_sign_frac_yangpeng_2024} while a comprehensive study of the effect of using various 5G NR reference signals for sensing was given in \cite{isac_vehicle_nataraja_2024}.

In this paper, we address a novel model-based estimation scheme that addresses the following two goals: i) to obtain high resolution estimates of \gls{icas} target parameters (bistatic delay and Doppler) from sparsely allocated \gls{ofdm} \gls{re} in time and frequency and ii) to solve the problem that arises from normalizing the received signal relative to the estimated transmit signal, which is used as a correlation reference at the receiver. The first problem arises because of OFDMA multiuser access and \gls{tdd} operation. The latter results from the multilevel \gls{qam} carrier modulation of the data payload waveform, which causes noise enhancement when not accounted for carefully. 

The method is based on an estimation framework proposed for wideband channel sounding \cite{hrpe_wideband_semper_2023}. It can be extended to a joint multidimensional estimation procedure that includes directional propagation parameters at transmit and receiver side depending on the availability of antenna arrays.

\section{Notation}
We use boldface letters to denote N-way arrays. The operator $\left[ \cdot \right]_{ij}$ is used to denote the element in the $i$th row and $j$th column of a matrix. The operators $\left( \cdot \right)^H$ and $\odot$ denote the complex-conjugate transpose and element-wise multiplication, respectively. Further, $\otimes$ is used to describe the outer-product of two 1-way arrays and $\mathrm{diag}(\cdot)$ to construct a diagonal matrix from the 1-way array operand.

\section{Propagation Model}
We consider \gls{ofdm} signals that contain structured reference signals and arbitrary user data drawn from a \gls{qam} modulation alphabet. We express the transmitted signal for a block of $N_T$ symbols with $N_C$ sub-carriers as
\begin{equation}
    x(t)=\sum_{n=0}^{N_T-1} \sum_{k=0}^{N_c-1} \left[\mathbf{X}\right]_{n,k} \exp(j 2 \pi k \Delta f t) u_{T}\left(t-n T_{\mathrm{O}}\right),
    \label{eq:ofdm_tx_all}
\end{equation}
where $\left[\mathbf{X}\right]_{n,k} \in \C$ denotes the complex data on $n$th symbol and $k$th sub-carrier, $\Delta f$ the sub-carrier spacing, $T_O$ the \gls{ofdm} symbol duration including cyclic prefix and guard interval and $u_T(t)$ is the rectangular function of length $T$.

In a \gls{tdd} system, both transmission and reception occur on the same frequency, resulting in non-continuous transmission. This introduces temporal gaps within the coherent processing block at the receiver, i.e. not all $N_T$ symbols carry data. Furthermore, as considered in \cite{wf_design_mura_2023}, we assume that only a subset of sub-carriers are activated for any \gls{ofdm} symbol, i.e. the \gls{ro} $\zeta < 1$. These two assumptions introduce sparsity in the symbol and sub-carrier domains respectively. 

We denote as $\mathcal{S}$ and $\mathcal{C}$ to be the sets of all sub-carrier and symbol indices in a coherent processing block, such that $\mathcal{D} \subseteq \mathcal{S} \times \mathcal{C} $ and $\mathcal{P} \subseteq \mathcal{S} \times \mathcal{C}$ represent the sets of symbol and sub-carrier index pairs $(s,c)$ utilized for data- and reference signal-bearing \glspl{re}, respectively. The transmitted \gls{ofdm} frame can then be expressed as
\begin{equation}
    \mathbf{X} = \eta\mathbf{X}_{\mathcal{D}} + \beta\mathbf{X}_{\mathcal{P}},
    \label{eq:ofdm_tx_pilot_data}
\end{equation}
where the terms $\eta, \beta \in [0, 1)$ denote the distribution of power between data and pilot \glspl{re} and $\mathbf{X}_{\mathcal{D}}$, $\mathbf{X}_{\mathcal{D}}$ denote the elements of $\mathbf{X}$ corresponding to pilot and data \glspl{re}. Furthermore, we point out that the summands in \cref{eq:ofdm_tx_pilot_data} are perfectly orthogonal to each other due to the use of \gls{ofdm}, i.e. $\mathcal{D} \cap \mathcal{P}$ is an empty set and that the elements of $\mathbf{X}_{\mathcal{D}}$ are drawn randomly from a modulation alphabet $\mathcal{M}$, representing an arbitrary stream of user data.

\begin{figure}[ht]
\centering
\begin{subfigure}[b]{0.23\textwidth}
    \centering
    \includegraphics[width=\textwidth]{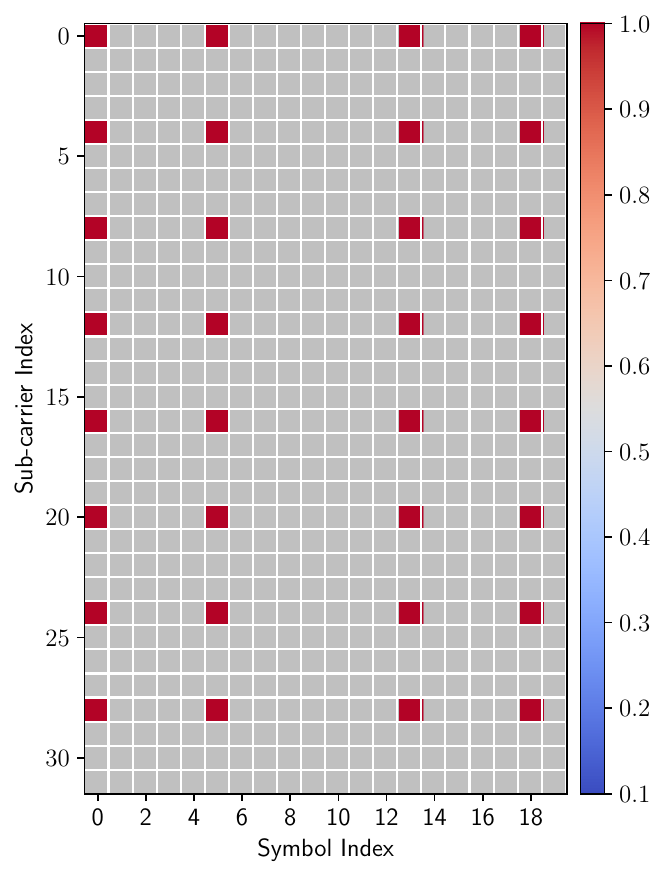}
    \caption{Only pilot \glspl{re}}
    \label{subfig:rg_only_pilots}
\end{subfigure}
\begin{subfigure}[b]{0.23\textwidth} 
    \centering
    \includegraphics[width=\textwidth]{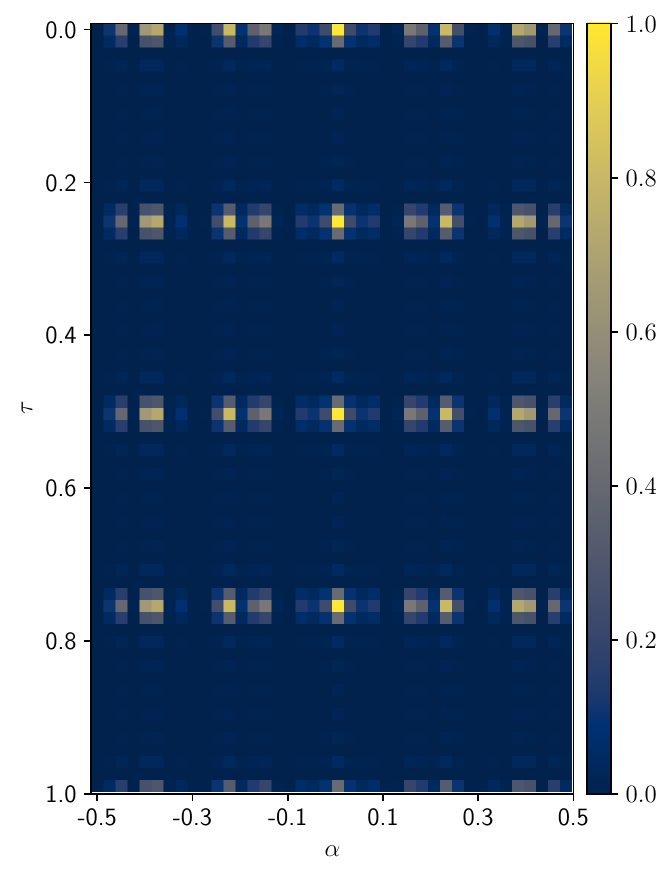}
    \caption{Resulting AF}
    \label{subfig:acf_only_pilots}
\end{subfigure}
\hfill

\begin{subfigure}[b]{0.235\textwidth}
    \centering
    \includegraphics[width=\textwidth]{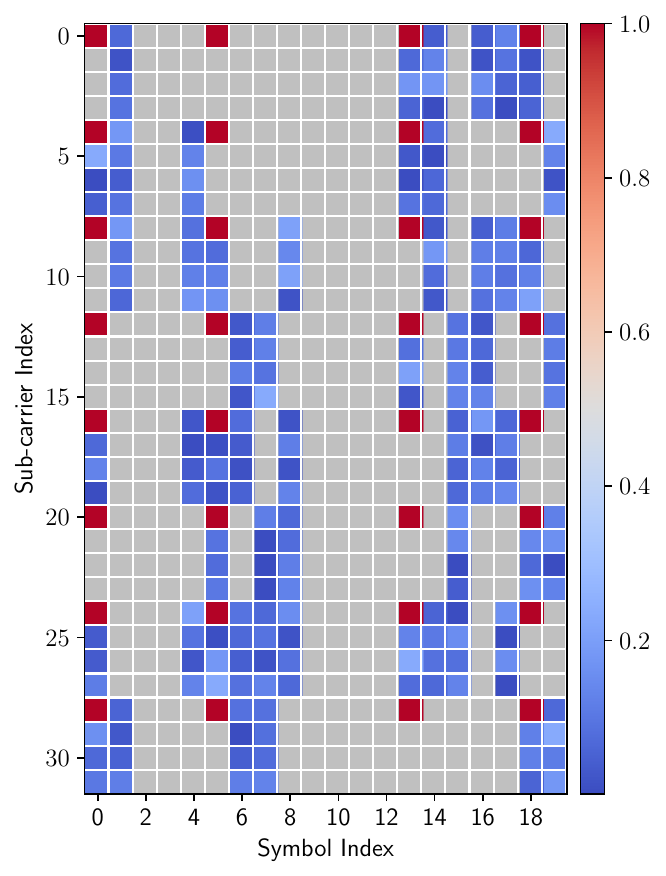}
    \caption{Pilot and data \glspl{re}}
    \label{subfig:rg_data_pilots}
\end{subfigure}
\begin{subfigure}[b]{0.235\textwidth}
    \centering
    \includegraphics[width=\textwidth]{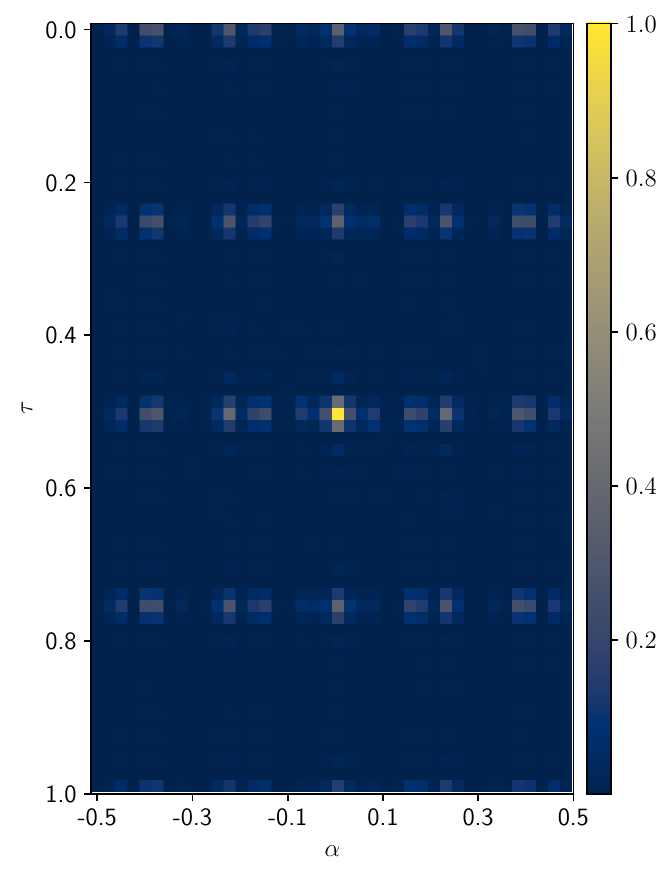}
    \caption{Resulting AF}
    \label{subfig:af_data_pilots}
\end{subfigure}

\caption{Illustrates the improvement in the \gls{af} attainable by utilizing data-bearing \glspl{re} in addition to pilot-bearing \glspl{re}. The red \glspl{re} in (a) and (c) denote the high relative power allocated to the pilot-bearing \glspl{re}. In (b) and (d) we see the normalized magnitude of the resulting \glspl{af}.}
\label{fig:resource_grid_af}
\end{figure}

The continuous time received signal at the receiver can be expressed as \cite{effects_arbitrarily_subcarr_fink_2012}
\begin{equation}
    y(t) = \sum_{p = 1}^{P} \gamma_p x(t - \tau_p) \exp(j2\pi \alpha_p t) + n(t),
    \label{eq:rx_signal_time}
\end{equation}
where $P$ is the total number of incident scattered paths (corresponding to one or more targets) and $n(t) \sim \mathcal{CN}\left(0, \delta \right)$ the complex additive Gaussian white noise. Each scattered path causes a shift in time delay $\tau_p \in \R$ and frequency $\alpha_p \in \R$ of the transmitted signal $x(t)$ and is attenuated and phase-shifted by the complex weight $\gamma_p \in \C$ that accounts for path loss and the reflectivity of scattering object(s). The time-domain signal expressed in \cref{eq:rx_signal_time} is sampled at the receiver and the base-band received signal matrix $\mathbf{Y} \in \C^{N_F \times N_T}$ is constructed via a $N_{\text{F}}$-point \gls{dft}. It can be expressed as
\begin{equation}
    \mathbf{Y} = \mathbf{X} \odot \mathbf{H} + \mathbf{N}.
    \label{eq:rx_signal_freq}
\end{equation}
We assume that sufficient synchronization is possible between the TX and RX and do not consider the effect of sampling time and frequency offsets \cite{bistatic_ofdm_jrc_de_oliveira_2023}.

Further, we assume that the receiver has a-priori knowledge of the positions of the utilized \glspl{re}, namely the union set $\mathcal{U} = \mathcal{P} \cup \mathcal{D} \subset \mathcal{S} \times \mathcal{C}$, or is able to gain this information using an appropriate procedure. An estimate of the transmitted data $\hat{\mathbf{X}}_{\mathcal{D}}$ is assumed to be available, either by exact knowledge (mono-static case) or via standard \gls{ofdm} decoding (bi-static case). For this work, we assume that perfect decoding is possible, i.e. $\hat{\mathbf{X}} = \mathbf{X}$ since the focus is to propose an estimator that can operate with arbitrary user data. In the rest of this work, the delay and Doppler-shift parameters, $\tau$ and $\alpha$, are normalized by the total bandwidth $N_{F} \Delta F$ \SI{}{Hz} and coherent block length $N_T T_0$ seconds, such that $\tau \in (0, 1]$ and $\alpha \in (-0.5, 0.5]$.

\begin{figure}[ht]
    \centering
    \begin{subfigure}[b]{0.24\textwidth}
        \centering
        \includegraphics[width=\textwidth]{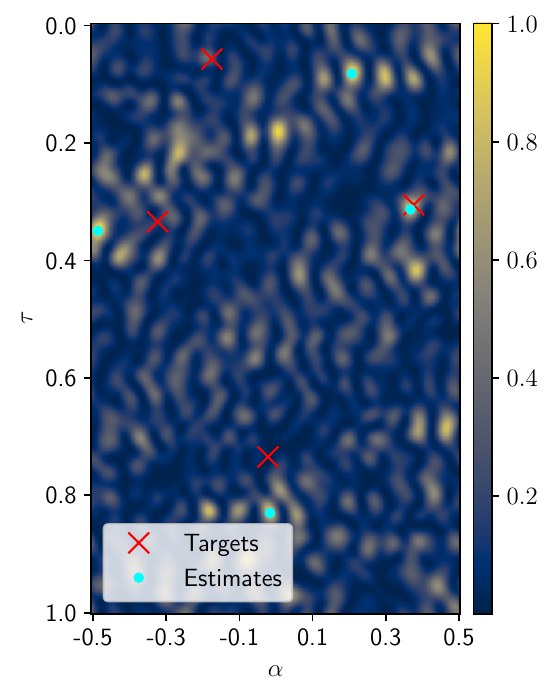}
        \caption{$|\mathbf{a}(\tau, \alpha)^H \hat{\mathbf{h}}|^2$}
        \label{subfig:sp_func_deconvolved}
    \end{subfigure}
    \begin{subfigure}[b]{0.24\textwidth} 
        \centering
        \includegraphics[width=\textwidth]{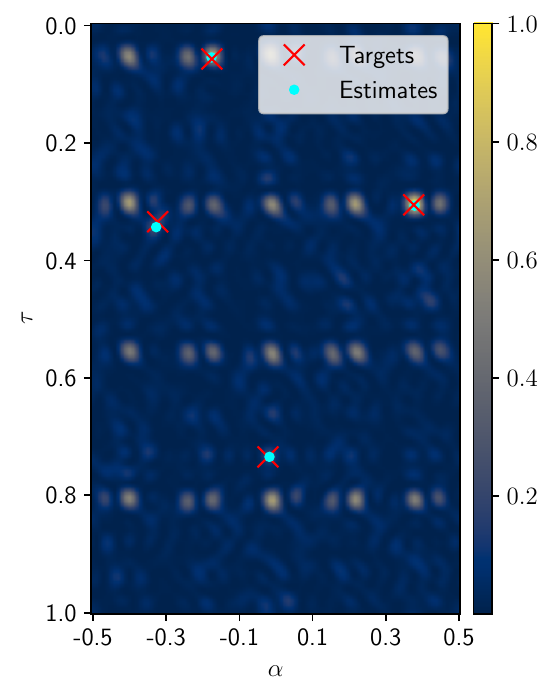}
        \caption{$|C(\tau, \alpha)|^2$}
        \label{subfig:sp_func_direct}
    \end{subfigure}
    
    \caption{Illustrates the normalized magnitude of the Spreading Function in a 4 path, low-\gls{snr} scenario calculated by: (a) correlating the Zero-Forcing channel estimate $\hat{\mathbf{h}}$ with a model of the channel response and (b) correlating the received signal with a weighted reference signal. Furthermore, target parameters and estimates obtained show that the estimator performs better using the latter correlation procedure.}
    \label{fig:spreading_funcs}
\end{figure}

\section{Resource Sparsity and Amplitude Modulation}\label{sec:sparsity_amplitude_mod}
We propose the utilization of data-bearing \glspl{re} in addition to reference signals for sensing due to the gain in \gls{snr} achievable in comparison to using only the latter. Furthermore, due to the sparse and structured nature of reference signal allocation in existing standards, the \gls{af} resulting from the usage of only pilot \glspl{re} exhibits strong aliasing side-lobes as opposed to when data \glspl{re} are additionally used (see \Cref{fig:resource_grid_af}).

A common first step in \gls{ofdm} radar estimation algorithms is to obtain a channel estimate $\hat{\mathbf{H}}$, which is then used to compute the magnitude of the \gls{sf} via a 2-D \gls{dft}. When considering signals drawn from a \gls{qam} scheme, obtaining $\hat{\mathbf{H}}$ via the \gls{zf} solution causes an amplification of noise for those \glspl{re} with small instantaneous magnitude. The \gls{mf} approach, where the channel estimate is computed via a cross-correlation $\mathbf{H} = \mathbf{X}^\ast \mathbf{Y}$, is more robust to noise but suffers from the appearance of side-lobes due to the use of \gls{qam}. These issues are further aggravated by the practice of reference signal boosting, where signals such as DM-RS are given a higher share of power than data payloads to facilitate better channel estimation for communication decoding.

Additional non-ideal side-lobes are introduced when performing \gls{dft}-based processing due to the sparsity of data-bearing \glspl{re} in time and frequency arising from low \gls{ro} and temporal gaps due to \gls{tdd} schemes. Proposed methods such as zero-filling and linear interpolation cause biased delay-Doppler estimates and degrade estimation performance while matrix interpolation via rank-reduction techniques \cite{wf_design_mura_2023} suffer in clutter-rich environments \cite{model_based_jdd_miranda_2025}.

To overcome the above-mentioned problems when performing sensing with sparse \glspl{re}, we propose the use of model-based estimation to model the channel response at only \emph{non-empty} \glspl{re}, avoiding the problems of zero-filling or incorrect interpolation \cite{model_based_jdd_miranda_2025}. Furthermore, we modify the device model used in the RIMAX \cite{rimax_thoma_2004} estimator to incorporate the estimated transmit signal $\hat{\mathbf{X}}$, hence accounting for the effect of arbitrary user payloads.

\section{Device Model} \label{sec:device_model}
To address the observed anomalies in the received signal arising from the assumptions outlined in \Cref{sec:sparsity_amplitude_mod}, the proposed estimation procedure utilizes a device model that incorporates the specific time-frequency sampling pattern and the instantaneous amplitude of the transmit signal's \glspl{re}. This facilitates a physically meaningful representation of the received signal at the non-empty \glspl{re} while avoiding the need for explicit interpolation.

Considering a complete coherent processing block sampled at frequencies $\mathbf{f} \in \R^{N_F}$ and time instances $\mathbf{t} \in \R^{N_T}$, the non-linear contribution of the delay and Doppler-shifts to the channel transfer function $\mathbf{H}$ is described by the array $\hat{\mathbf{a}}: \R \times \R \rightarrow \C^{N_F \times N_T}$ defined as
\begin{align}
    \hat{\mathbf{a}}(\tau, \alpha) &= \exp(-j2\pi \mathbf{f}\tau) \otimes \exp(j2\pi \mathbf{t}\alpha).
\end{align}
To model the channel response at only the useful non-empty \glspl{re} indexed by the set $\mathcal{U}$, we stack the elements in a 1-way array $\mathbf{a}: \R \times \R \rightarrow \C^{N_{\mathcal{U}}}$ where
\begin{equation}
    \mathbf{a}(\tau, \alpha) = \vecop\left( \hat{\mathbf{a}}_{\mathcal{U}} \right),
    \label{eq:vec_a}
\end{equation}
and $N_{\mathcal{U}} = \left\vert \mathcal{U} \right\vert$ denotes the number of utilized \glspl{re} (data and reference) in the complete coherent processing block.
For $P$ propagating paths, the expected noise-less samples of the channel transfer function are described by
\begin{equation}
    \mathbf{h}(\bm{\tau}, \bm{\alpha}, \bm{\gamma}) = \sum_{p = 1}^{P} \left[\bm{\gamma}\right]_p \mathbf{a}\left(\left[\bm{\tau}\right]_p, \left[\bm{\alpha}\right]_p \right)
    \label{eq:H_matrix}
\end{equation}
where $\bm{\tau}, \bm{\alpha} \in \R^P$ and $\bm{\gamma} \in \C^P$ contain the delay, Doppler-shift and complex path weight parameters for each path.
  
Furthermore, to account for the instantaneous magnitude of the \glspl{re} in the transmit signal, we model the noise-less received signal at indices $\mathcal{U}$ as $\mathbf{s}: \R^P \times \R^P \times \C^P \times C^{N_{\mathcal{U}}} \rightarrow \C^{N_\mathcal{U}}$ defined as
\begin{equation}
    \mathbf{s}(\bm{\tau}, \bm{\alpha}, \bm{\gamma}, \hat{\mathbf{X}}) = \vecop(\hat{\mathbf{X}}_{\mathcal{U}}) \odot \mathbf{h}_{\mathcal{U}}(\bm{\tau}, \bm{\alpha}, \bm{\gamma}),
    \label{eq:weighted_model}
\end{equation}
where $\hat{\mathbf{X}}$ is an estimate of the transmit signal. For brevity, we drop the term $\hat{\mathbf{X}}$ in further descriptions of \cref{eq:weighted_model}.
We see that the model in \cref{eq:weighted_model} weights the elements of the expected 2-D complex exponentials that arise due to propagation path scattering, described in \cref{eq:H_matrix}. In the next section, we describe how this device model is used for target parameter estimation.

\section{Delay-Doppler Estimation using OFDMA Signals} \label{sec:estimation_proced}
Similar to \cref{eq:vec_a}, the elements of the frequency domain received signal from \cref{eq:rx_signal_freq} are stacked in the array $\mathbf{y} = \vecop(\mathbf{Y}_{\mathcal{U}}) \in \C^{N_\mathcal{U}}$. Estimates of the delay and Doppler-shift $\hat{\bm{\tau}}$ and $\hat{\bm{\alpha}}$ parameters are then obtained by solving the problem
\begin{equation}
    \arg\min_{
        \bm{\tau}, \bm{\alpha}
    } \left( 
        \mathbf{y} 
        - \mathbf{s}(
            \bm{\tau}, \bm{\alpha}, \bm{\gamma}
        ) 
    \right)^{\text{H}} 
    \mathbf{R}^{-1} 
    \left(
        \mathbf{y} 
        - \mathbf{s}(
            \bm{\tau}, \bm{\alpha}, \bm{\gamma}
        ) 
    \right),
    \label{eq:ml_cost_function}
\end{equation}
where $\mathbf{R} \in \R^{N_{\mathcal{U}} \times N_{\mathcal{U}}}$ denotes the noise covariance. As we assume un-correlated additive Gaussian white noise, the $\mathbf{R} = \sigma^2 \mathbf{I}$. Since \cref{eq:ml_cost_function} is non-convex \cite{estimation_radio_params_richter_2005}, we utilize an iterative estimation procedure as follows: 
\begin{itemize}
\item coarse estimation of the delay and Doppler-shift parameters of a single path via a peak-search of the \gls{sf}
\item calculating $\gamma \in \C$ via a least-squares solution which constitutes the \gls{blue}
\item refining the estimates using Fischer-scored gradient iterations
\item subtracting the complex contribution of the estimated path from the received signal $\mathbf{y}$ to get the residual \cref{eq:residual}
\item repeating the procedure above for multi-path estimation.
\end{itemize}

\subsection{Coarse Estimation}\label{subsec:coarse_estimation}
We perform a peak-search of the magnitude of the \gls{sf} $C: \R \times \R \rightarrow \C$ described as
\begin{equation}
    C(\tau, \alpha) = \left( \vecop(\hat{\mathbf{X}}_{\mathcal{U}}) \odot \mathbf{a}(\tau, \alpha) \right)^H \mathbf{y},
    \label{eq:weighted_spreading_func}
\end{equation}
to obtain coarse estimates of the target delay and Doppler shift parameters. In \Cref{subfig:sp_func_direct}, we see that the intermediate \gls{sf} obtained using \cref{eq:weighted_spreading_func} is more robust to instantaneous noise and results in improved target estimates compared to \Cref{subfig:sp_func_deconvolved}, where a \gls{zf} estimate of $\hat{\mathbf{h}}$ is correlated with the \emph{non-weighted} device model from \cref{eq:vec_a}. 

Note, that if \cref{eq:weighted_spreading_func} is computed using the \gls{dft}, this step is equivalent to \gls{mf} response used in radar algorithms. However, in this case \gls{dft} can be used only if the \glspl{re} are allocated sufficiently randomly \cite{model_based_jdd_miranda_2025}.

\begin{figure}[t]
    \centering
    \includegraphics[width=\columnwidth]{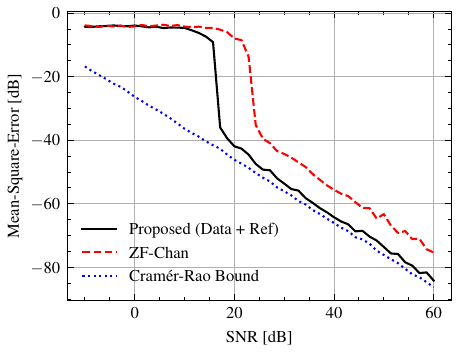}
    
    \caption{Plots the Mean-Squared estimation error for the delay and Doppler-shift parameters along-side the lower bound on estimate variance.}
    \label{fig:mse}
\end{figure}

\subsection{Path-weight Estimation}
For a given set of $p < P$ previously estimated delay and Doppler-shift parameters $\hat{\bm{\tau}}^p$ and $\hat{\bm{\alpha}}^p$, we define the matrix valued function $\mathbf{Z}: \R^p \times \R^p \rightarrow \C^{N_{\mathcal{U}} \times p}$ as
\begin{equation}
    \mathbf{Z}(\hat{\bm{\tau}}^p, \hat{\bm{\alpha}}^p) = \mathrm{diag}(\hat{\mathbf{X}}_{\mathcal{U}}) \mathbf{A}(\hat{\bm{\tau}}^p, \hat{\bm{\alpha}}^p),
    \label{eq:blue_intermed}
\end{equation}
where the columns of $\mathbf{A}: \R^p \times \R^p \rightarrow \C^{N_{\mathcal{U}} \times p}$ correspond to the individual complex response arrays of each path.
An estimate of the linear path-weights $\hat{\bm{\gamma}} \in \C^p$ for all $p$ paths are then jointly estimated via a \gls{wls} operation, i.e.
\begin{equation}
    \hat{\bm{\gamma}}^p = \left( \mathbf{Z}(\hat{\bm{\tau}}^p, \hat{\bm{\alpha}}^p)^H \mathbf{R}^{-1} \mathbf{Z}(\hat{\bm{\tau}}^p, \hat{\bm{\alpha}}^p) \right)^{-1} \mathbf{Z}(\hat{\bm{\tau}}^p, \hat{\bm{\alpha}}^p)^H \mathbf{R}^{-1} \mathbf{y},
    \label{eq:blue_estim}
\end{equation}
which describes the \gls{blue} of $\bm{\gamma}$. We see from \cref{eq:blue_estim,eq:blue_intermed} that computation of $\gamma$ is equivalent to a \gls{wls} operation due to the inclusion of the instantaneous magnitude of each \gls{re} in the transmit signal.

\subsection{Gradient-based Optimization}
To obtain off-grid estimates, the coarse parameters $\hat{\bm{\tau}}_p$ and $\hat{\bm{\alpha}}_p$ are further refined using the second-order Levenberg-Marquardt \cite{levenberg_marquadt_1944} method, formulations for which are given in \cite[Section 5.2.4]{estimation_radio_params_richter_2005}. The number of gradient updates to the coarse estimates can be chosen such that the required balance between computational time and obtained estimate error is achieved \cite{model_based_jdd_miranda_2025}.

\subsection{Multi-path Estimation}
To estimate multiple paths, the contribution of previously estimated paths are iteratively subtracted from the received signal, i.e. for every path iteration $p < P$, the current residual is computed as
\begin{equation}
    \mathbf{r}^p \left(\bm{\tau}^p, \bm{\alpha}^p, \bm{\gamma}^p \right) = \mathbf{y} - \mathbf{s}(\bm{\tau}^p, \bm{\alpha}^p, \bm{\gamma}^p),
    \label{eq:residual}
\end{equation} 
where $\bm{\tau}^p, \bm{\alpha}^p, \bm{\gamma}^p$ denote the previously estimated parameters of $p$ paths. Due to the use of a physically justified device model \eqref{eq:H_matrix}, the contribution of each path includes all non-ideal effects resulting from the assumptions outlined in \Cref{sec:sparsity_amplitude_mod} and is effectively canceled before estimating the next path $p+1$ from the residual $\mathbf{r}^p$. While the model order can be estimated by monitoring and thresholding the values of the residual from \cref{eq:residual}, we assume a known model order for this work.

\begin{figure}[ht]
    \centering
    \begin{subfigure}{\columnwidth}
        \centering
        \includegraphics[width=0.9\textwidth]{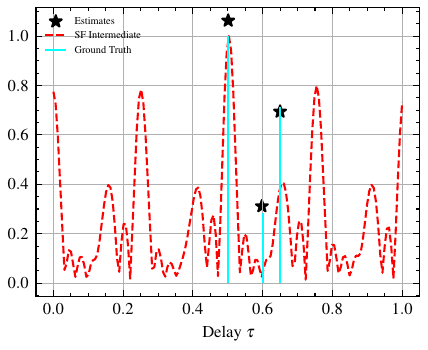}
        \caption{Normalized Magnitude of intermediate \gls{sf} as in \eqref{eq:weighted_spreading_func}}
        \label{subfig:sf_1d_proposed}
    \end{subfigure}
    
    \vspace{0.2cm} 
    
    \begin{subfigure}{\columnwidth}
        \centering
        \includegraphics[width=0.9\textwidth]{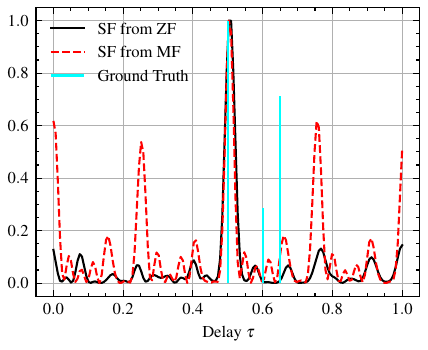}
        \caption{Normalized magnitudes of \glspl{sf} obtained via \gls{zf} and \gls{mf}}
        \label{subfig:sf_1d_trad}
    \end{subfigure}
    
    \caption{The figure in (a) plots the normalized magnitude of the intermediate \gls{sf} used for coarse estimation in the proposed estimator for a 1-D slice along the delay axis. In (b), the \glspl{sf} computed using the \gls{zf} and \gls{mf} followed by \gls{dft} are processing are compared. In both plots the SNR is set to $10$ \unit{dB}. We see in (a) that the estimator is able to correctly estimate the delay parameter and path-weights even in the presence of side-lobes.}
    \label{fig:sf_1d}
\end{figure}

\section{Numerical Verification}
For the purpose of conducting Monte Carlo simulations, we adopt system parameters of $N_F = 32, N_T = 20$, a resource occupancy factor of $\eta = 0.4$ and 256-QAM modulation. Although these specific values are not explicitly defined in the 5G-NR standard, they are representative of practical scenarios, and the presented results remain applicable across different numerologies.

In \Cref{fig:mse}, we consider $P = 3$ paths with delay and Doppler-shift parameters drawn randomly every simulation run. We plot the \gls{mse} of the estimates for two estimation procedures, namely
\begin{itemize}
    \item The proposed estimator using pilots and data \glspl{re} following the procedure in \Cref{sec:estimation_proced} and using the device model defined in \Cref{sec:device_model},
    \item Performing delay-Doppler estimation after obtaining a \gls{zf} channel estimate $\hat{\mathbf{h}}$ using the procedure defined in \Cref{sec:estimation_proced} without the incorporation of the transmit signal in the device model.
\end{itemize}

To understand how the coarse estimation step in \Cref{subsec:coarse_estimation} compares to the computation of the \gls{sf} in radar processing algorithms, we plot in \Cref{fig:sf_1d} the intermediate \gls{sf} as described in \cref{eq:weighted_spreading_func} and compare it to the \gls{sf} obtained via \gls{dft} processing of the channel estimate obtained via
\begin{itemize}
    \item the \gls{mf} approach where $[\hat{\mathbf{h}}]_i = [\mathbf{X}]_i^* [\mathbf{y}]_i$ and
    \item the \gls{zf} approach where $[\hat{\mathbf{h}}]_i = [\mathbf{y}]_i / [\mathbf{X}]_i$.
\end{itemize}
followed by zero-filling for the empty \glspl{re}.
We see in \Cref{subfig:sf_1d_proposed} that the estimator is able to correctly estimate the 3 path parameters even in the presence of sidelobes in \cref{eq:weighted_model}. Although counterintuitive, this can be explained by the successive path cancellation step in \cref{eq:residual}, which subtracts each individual path \emph{including its side-lobes} from the received signal.

\section{Conclusions and Future Work}
A further study on the effect of decoding errors on sensing performance in bi-static setups is planned. Furthermore, the estimation procedure proposed in this work will be validated on real-world \gls{icas} channel measurements followed by investigations into run-time complexity and computational requirements. While this work focused on estimation techniques, a practical system must also consider wave-form design for to improve the sensing \gls{af}.

\section*{Acknowledgment}
This work is partly sponsored by the \textit{Deutsche Forschungsgesellschaft (DFG)} under the research projects JCRS CoMP with Grant-No. TH $494/35-1$ and by the BMBF project 6G-ICAS4Mobility with Project No. 16KISK241.

\textbf{Disclaimer}: This paper has been submitted to the European Conference on Networks and Communications (EUCNC) 2025. The content may undergo revisions based on peer-review feedback.

\printbibliography
\end{document}

%% file: abbrevs.tex
\newacronym{aoa}{AoA}{Angle-of-Arrival}
\newacronym{aod}{AoD}{Angle-of-Departure}
\newacronym{adam}{ADAM}{ADAM}
\newacronym{af}{AF}{Ambiguity Function}
\newacronym{blue}{BLUE}{Best-Linear-Unbiased Estimate}
\newacronym{cpcl}{CPCL}{Cooperative Passive Coherent Location}
\newacronym{crb}{CRB}{Cramér-Rao Lower Bound}
\newacronym{cs}{CS}{Compressed Sensing}
\newacronym{dft}{DFT}{Discrete Fourier-Transform}

\newacronym{eadf}{EADF}{Effective Aperture Distribution Function}

\newacronym{ft}{FT}{Frequency-time}

\newacronym{icas}{ICAS}{Integrated Communications and Sensing}

\newacronym{ir}{IR}{Impulse Response}
\newacronym{mf}{MF}{Matched filtering}
\newacronym{ml}{ML}{Maximum-Likelihood}
\newacronym{mmwave}{mmWave}{Millimeter-wave}
\newacronym{mse}{MSE}{Mean-Square Error}

\newacronym{ofdm}{OFDM}{Orthogonal frequency-division multiplexing}
\newacronym{ofdma}{OFDMA}{Orthogonal Frequency-division Multiple Access}

\newacronym{prs}{PRS}{Positioning Reference Symbols}

\newacronym{qam}{QAM}{Quadrature Amplitude Modulation}

\newacronym{re}{RE}{Resource Element}

\newacronym{ro}{RO}{Resource Occupancy}

\newacronym{rmse}{RMSE}{Root-Mean-Square Error}

\newacronym{sf}{SF}{Spreading function}
\newacronym{scf}{SCF}{Scattering Function}
\newacronym{sgd}{SGD}{Stochastic Gradient Descent}
\newacronym{snr}{SNR}{Signal-to-noise Ratio}

\newacronym{tdd}{TDD}{Time-Division-Duplex}

\newacronym{uav}{UAV}{Unmanned Aerial Vehicle}

\newacronym{wls}{WLS}{Weighted Least-Squares}

\newacronym{xr}{XR}{Extended Reality}

\newacronym{zf}{ZF}{Zero-Forcing}

%% file: macros.tex
\newcommand{\R}{{\mathbb R}}
\newcommand{\C}{{\mathbb C}}

\DeclareMathOperator{\vecop}{\mathrm{vec}}